\documentclass[11pt]{revtex4-1}
\usepackage{graphicx}
\usepackage{amsmath}

\begin{document}

%%%%%%%%%%%%%%%%%% title page information %%%%%%%%%%%%%%%%%%
\title{Simultaneous measurements of electrophoretic and dielectrophoretic forces using 
optical tweezers }
\author{Giuseppe Pesce}
\email{giuseppe.pesce@fisica.unina.it} 
\affiliation{Dipartimento di Fisica, Universit\`a degli studi di Napoli ''Federico II'' \\
Complesso Universitario Monte S.Angelo, Via Cintia, 80126 Napoli, Italy}

\author{Giulia Rusciano}
\affiliation{Dipartimento di Fisica, Universit\`a degli studi di Napoli ''Federico II'' \\
Complesso Universitario Monte S.Angelo, Via Cintia, 80126 Napoli, Italy}

\author{Gianluigi Zito}
\affiliation{Dipartimento di Fisica, Universit\`a degli studi di Napoli ''Federico II'' \\
Complesso Universitario Monte S.Angelo, Via Cintia, 80126 Napoli, Italy}

\author{Antonio Sasso}
\affiliation{Dipartimento di Fisica, Universit\`a degli studi di Napoli ''Federico II'' \\
Complesso Universitario Monte S.Angelo, Via Cintia, 80126 Napoli, Italy}

\begin{abstract}
Herein, charged microbeads handled with optical tweezers are used as a sensitive probe for simultaneous 
measurements of electrophoretic and dielectrophoretic forces. We first determine the electric charge carried 
by a single bead by keeping it in a predictable uniform electric field produced by two parallel planar 
electrodes, then, we examine same bead's response in proximity to a tip electrode. In this case, besides electric forces, the bead simultaneously experiences non-negligible dielectrophoretic forces produced by the strong electric field gradient.  The stochastic and deterministic motions of the trapped bead are theoretically and 
experimentally analysed in terms of the autocorrelation function. By fitting the experimental data, we are 
able to extract simultaneously the spatial distribution of electrophoretic and dielectrophoretic forces around 
the tip. Our approach can be used for determining actual, total force components in the presence of high-curvature electrodes or metal scanning probe tips.
\end{abstract}

\maketitle
%\ocis{(140.7010) Laser trapping; (170.4520) Optical confinement and manipulation; (350.4855) Optical tweezers or optical manipulation.}

%%%%%%%%%%%%%%%%%%%%%%% References %%%%%%%%%%%%%%%%%%%%%%%%%

%\bibliography{depbib}{}
%\bibliographystyle{osajnl}

%%%%%%%%%%%%%%%%%%%%%%%%%%  body  %%%%%%%%%%%%%%%%%%%%%%%%%%
\section{Introduction}
The behavior of charged and neutral microparticles in uniform and nonuniform electric fields is a well established issue \cite{Pohl1,Pohl2}. Nevertheless, in recent years, interest has grown in application of electrophoretic (EP) and  dielectrophoretic (DEP) forces in a wide variety of micro-systems. In particular, EP and DEP forces are revealing themselves as a useful tool in biology and medicine for precise positioning and manipulation of single cells or bacteria. In addition, specific field configurations with a high electric field gradient can be designed  with the aid of micromachined electrodes allowing one to achieve actual lab-on-a-chip tweezer devices. Micro-objects have been efficiently trapped  \cite{Kodama,Cheng,Kua}, and many biological applications have also been proposed
(see \cite{Pethig} for a review) such as, for instance, virus trapping in high-frequency electric field cages \cite{Schnelle}. In addition, DEP forces have been used in microfluidics devices for single- and multi-cell sample preparation \cite{Hunt,Gray,IliescuAPL07} or to achieve lateral deviation of particles in liquid flows \cite{Demierre}. DEP tweezers have been also realized by Lee \textit{et al.} \cite{Lee} in a fluid environment by means of a localized 3D movable electric field configuration. 

Optical Tweezers (OT) \cite{AshkinOL86} are formidable and versatile tools for manipulating  dielectric particles or cells  having size ranging from tens of nanometers up to tens of microns and exerting forces ranging from 10 fN to 100 pN.  Therefore, it is not surprising that EP, DEP and photonic forces were combined to compare the range of applicability of the two trapping mechanisms, for reciprocal calibrations \cite{Fuhr,Schnelle1}, to evaluate trapping efficiencies \cite{Papagiakoumou}, or even to study molecular motors \cite{Arsenault}.
Typically, DEP forces are estimated by using the method of forced oscillation. In this case, an optically trapped bead is driven by a sinusoidal electric field; the particle displacement is monitored by a  position sensor and analyzed with a phase-sensitive detection scheme\cite{WeiB09,Hong,Zhu}. 

Recently, several experiments demonstrated the possibility to combine photonic and EP forces for estimating the electric charge carried by a single polystyrene microsphere \cite{GalnederBJ01,SethJCP07,SemenovJCIS09,BeunisPRL12,PesceE13,PesceCSB14}, or even for mapping the electric field generated in simple electrodes' geometries \cite{PesceLOC11}.
In the present work, we discuss a new approach to determine simultaneously EP and DEP forces in  proximity of microelectrodes. With respect to other approaches \cite{WeiB09}, our technique  is based on the analysis of the deterministic and stochastic motions of a trapped particle in terms of the autocorrelation function, and is able to provide, simultaneously, quantitative and absolute measurements of the particle charge, as well as EP and DEP forces.

\section{Theory}

A charged particle embedded in a fluid and confined in an optical trap, in presence of a nonuniform (\textit{i.e.}, its gradient $\nabla E\neq 0$) oscillating  electric field $E$,  undergoes five forces: (i) the elastic force of the trap, $F_{el}$; (ii) the drag force (Stokes force), $F_{drag}$; (iii) the electrophoretic force, $F_{EP}=QE$; (iv) the dielectrophoretic force, $F_{DEP}$; and (v) the stochastic thermal force. In the following, we will consider a one dimensional geometry for simplicity, however the results can be straightforwardly extended to three dimensional geometries. The particle trajectory  $x(t)$ of a microsphere of mass $m$, charge $Q$ and radius $a$, confined into an optical trap of stiffness $\kappa$ and embedded in a fluid of viscosity $\eta$ is ruled by the following Langevin equation:
\begin{align}
m \ddot{x} & =-\kappa x-\gamma \dot{x} +QE + 2\pi r^3 \varepsilon_{0} \varepsilon_{m} \mathcal{K}_{Re}(f_m)\nabla E^2 + \sqrt{2D}\xi(t),
\label{eq:langevin}
\end{align}
where $D=k_BT/\gamma$ is the diffusion coefficient, $\gamma=6\pi\eta a$ is the hydrodynamic factor, $k_B$ is the Boltzmann constant, $T$ is the absolute temperature, and $\xi(t)$ is a white noise term with zero mean, almost everywhere discontinuous and with infinite variation.   The fourth term in Eq. (\ref{eq:langevin}) represents the DEP force, where $ \mathcal{K}_{Re}(f_m)$ is the real part of the Claussius-Mossotti (CM) frequency-dependent function and $\varepsilon_{m}$ is the relative permittivity of water ($\sim$80).

For micrometric particles in water (low Reynolds number regime) the inertia is negligible, and by assuming that the particle is driven by a sinusoidal electric field $E(t)=E_0sin(2\pi f_m t)$, the solution of 
Eq. (\ref{eq:langevin}) is  given by:

\begin{flalign}
x(t) & =x_{th}+  \frac{Q\cdot E_0}{\kappa\sqrt{ 1+(f_m/f_C)^2}}sin(2\pi f_m t-\phi_{EP})  \\
 & + \frac{\pi r^3\varepsilon_m \varepsilon_0 \mathcal{K}_{Re}(f_m) \nabla E^2}{\kappa} \left[ 
1-\frac{ cos(4\pi f_m t-\phi_{DEP})}{ \sqrt{ 1+(2 f_m/f_C)^2}} \right]  =  \nonumber \\
&=x_{th} + Asin(2\pi f_m t-\phi_{EP})+B \left[ 
1-\frac{ cos(4\pi f_m t-\phi_{DEP})}{ \Delta} \right] ,
\label{eq:traj}
\end{flalign}
\noindent
where $x_{th}(t)$ is the thermal motion, while the second and third terms correspond, respectively, to the solutions of $EP$ and $DEP$ forces. The two phase terms that appear in Eq. (\ref{eq:traj})  are  defined as $tan(\phi_{EP})=f_m/f_C$ and  $tan(\phi_{DEP})=2f_m/f_C$. The {\it corner frequency} $f_C$  is related to the trap stiffness and to the drag coefficient by the relation $f_C=\kappa/(2\pi \gamma)$. It is worth noting that the DEP solution is composed by two terms:  a displacement of the bead from its equilibrium position in the optical trap and an oscillation at a frequency twice the modulation frequency of the forcing electric field. It is quite easy to calculate the autocorrelation function (acf) $\mathcal{C}(\tau)$:
\begin{align}
\mathcal{C}(\tau) &=\langle x(t)x(t+\tau) \rangle = \frac{k_B T }{\kappa}e^{-\tau/\tau_C} + 
 \frac{A^2}{2}cos(2\pi f_m \tau)+B^2 \left[ 1+\frac{ cos(4\pi f_m \tau)}{ 2\Delta^2} \right].
\label{eq:acf}
\end{align}
The decay time $\tau_C$   represents the characteristic time of the optical trap and is connected to the trap stiffness and hydrodynamic factor by the relation: $\tau_C=\gamma/\kappa=1/(2\pi \gamma)$ .

\section{Results and discussion}
The experimental setup is based on a custom-built optical microscope described  in \cite{PesceE13}. 
The laser beam (Nd-YAG laser, Innolight Mephisto NE500, $\lambda$= 1064 nm, 
maximum output power = 500 mW) was tightly focused with a high-numerical-aperture, water-
immersion objective lens (Olympus, UPLAPO60XW3, NA=1.2). Particle displacements were
measured by using a InGaAs Quadrant Photodiode (QPD, Hamamatsu G6849) at the back 
focal plane of the condenser lens \cite{GittesOL98}. 
Calibrations of trap and quadrant photodiode (QPD) were carried out using the well established power spectral density method \cite{SorensenRSI04,BuoscioloOC04}. In our experiment, the laser power was kept at $\sim$3 mW, which corresponds to a trap stiffness of about $1\times10^{-5}$ N/m and a characteristic time  $\tau_C \sim1$ ms.

\begin{figure}[t]
\begin{center}
\includegraphics[width=0.95\linewidth]{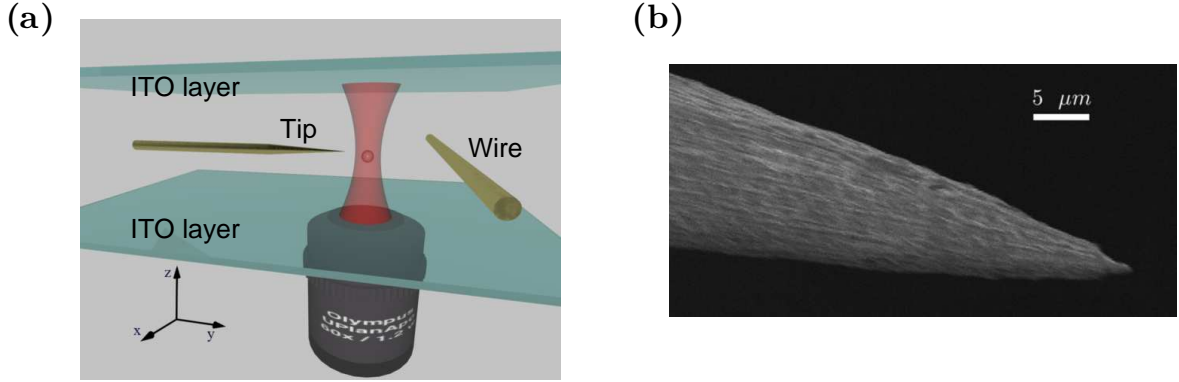}
\caption{(a) Sketch of the optically trapped, charged microsphere between planar, ITO covered (parallel) electrodes. A second set of electrodes (a wire and a tip) was placed in the same cell. The two parallel electrodes are employed to determine the charge carried by the bead. Instead, the wire and tip electrodes are used to analyze the EP and DEP forces. (b) SEM image of the tip.}
\label{fig:setup}
\end{center}
\end{figure}

Negatively charged, sulfate-coated  microspheres of polystyrene (Postnova, 1.06 g/cm$^3$ density, 1.65 refractive index) with a diameter of $1.00 \pm 0.05$ $\mu$m were diluted in distilled deionized water (conductivity $\sigma \sim1\,\mu S/cm$) to a final concentration of a few particles per microliter.  A droplet of such solution (50 $\mu$l) was injected inside a sandwiched chamber consisting of a 150 $\mu$m-thick coverslip and a microscope slide. Both  glass plates were  coated with an indium tin oxide (ITO) layer, hence forming parallel plate electrodes (Fig.\ref{fig:setup}(a)) that produced a uniform and predictable electric field $E= V_0/d$ at an applied voltage $V_0$ and electrode separation $d$. 
In the same cell, we also placed other two electrodes:  a sharp tip with radius of curvature $R\sim$150 nm and a gold coated tungsten wire ($\phi= 50~\mu$m)  perpendicular to each other (see Fig.\ref{fig:setup}(b)), separated by about 800 $mu$m. They were positioned in the middle of the sample cell (about 65 $\mu$m from the bottom coverslip) to avoid electro-osmotic flow effects\cite{SethJCP07}.

We measured the electric charge of a trapped bead applying the external voltage only at the parallel ITO electrodes as described in \cite{PesceE13}. In this case, the electric field was uniform ($B=0$ in Eq. (\ref{eq:traj}) and (\ref{eq:acf}), {\it i.e.}, no DEP forces were involved) and a trapped bead moved back and forth along the direction of the electric field ($z$-axis). Due to the finite drift velocity of free ions, the oscillating amplitude depends on the frequency of the electric field. We found the modulation frequency $f_m$ = 86.7 Hz to optimize the motion amplitude condition (for more details, see ref. \cite{PesceE13,PesceCSB14}). The bead's effective charge value resulted to be $Q=(-1.63 \pm 0.05)\times 10^{-16}~\mbox{C}$. At the end of the experiment, we measured again the charge and verified that its value was unchanged. It is worth noticing that  the measured charge resulted about three orders of magnitude lower than the value provided by the manufacturer ($Q=-1.79\times 10^{-13}~\mbox{C}$), but this discrepancy is ascribable to the screening effects caused by the free ions in water. Since the used polystyrene beads  showed certain degree of charge and radius heterogeneity, for accurate and reliable simultaneous force measurements, the second DEP experiment was carried out on the same trapped bead previously analyzed for charge measurement. 

Following the electric charge determination, we proceeded to measure the force field near the tip. In particular, we tracked the trajectories of a trapped microsphere in a raster grid around the tip. The sample cell was translated in the $x-y$ plane with steps of 2 $\mu$m in both directions as in scanning probe measurements. At each position of the grid, the particle trajectory was acquired for 20 s. All measurements were carried out by using a sinusoidal voltage amplitude $V_0 = 0.4$ V (between tip and wire) at a modulation frequency of 86.7 Hz. 
We chose an applied voltage small enough to avoid nonlinear phenomena in the fluid. At same time, this allowed to acquire bead displacements in a volume much smaller than the length scale  of variation of the resulting electric field gradient, therefore enabling a more accurate probing. We estimated the coefficients $A$ and $B$ of Eq. (\ref{eq:traj}), at a given position, from the amplitudes of the acf fuction. At the closest position to the tip where the gradient is larger, we found $A \approx 15~\mbox{nm and }B/\Delta \approx7~\mbox{nm}$. Such values confirm that in our experiment (i) the bead's excursion from the center of the optical trap results always small enough to fall in the linear range of our QPD ($\sim$300 nm in  our case) and that (ii) the bead's displacement is also much smaller than its radius. The first result guarantees that the bead trajectory is correctly sampled, whereas the latter shows that the bead oscillates in a very confined volume where the electric field and  its gradient can be reasonably assumed to be constant.  

\begin{figure}[t]
\begin{center}
\includegraphics[width=0.8\linewidth]{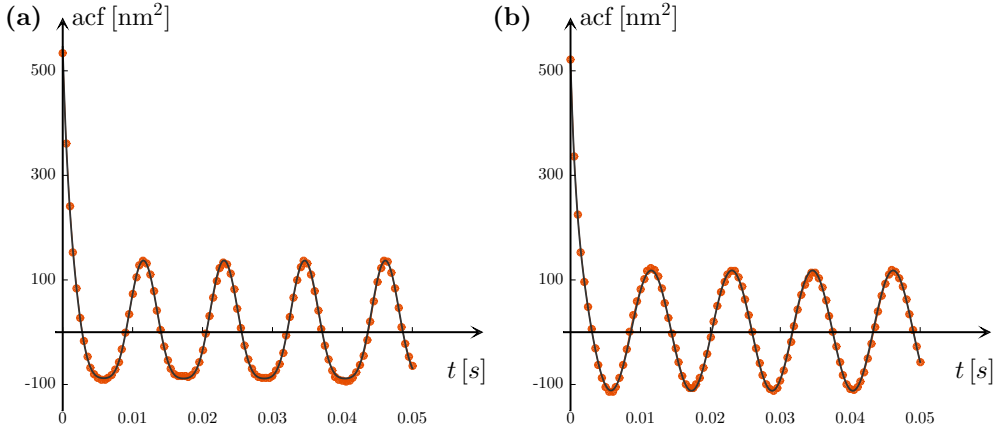}
\caption{Experimental acfs (dots)  and fitting to Eq. (\ref{eq:acf}) (solid lines) measured (a) when the bead is trapped near the tip, and (b) far away from the tip. Error bars are smaller than the markers size, the fitting produced a value of the reduced $\chi^2$ of 0.34 (panel (a)) and 0.47 (panel (b)), indicating its goodness}
\label{fig:acf}
\end{center}
\end{figure}

Figure \ref{fig:acf} shows the experimental acf curves along the $y$-axis, obtained by keeping the trapped bead at the two limit positions: the closest one to the tip (panel (a)), and a position far from the tip (panel (b)). We can observe that, while the oscillation is symmetric far away from the tip, it is instead clearly asymmetric in its proximity. This suggests that far from the tip, where the  electric field is essentially uniform, the bead motion is governed by the EP force only, {\it i.e.}, the acf is reduced to an oscillating term at frequency $f_m$ with amplitude $A^2/2$. On the contrary, near the tip where  $\nabla E \ne 0$,  the acf  contains also a second oscillating term at frequency 2$f_m$ with amplitude $B^2/2\Delta^2$.  Fitting our experimental data to Eq. (\ref{eq:acf}) as model, we found a very good agreement between theory and experiment as shown in Fig. \ref{fig:acf}. As discussed in Ref.\cite{PesceE13}, the same analysis can be performed using the power spectral density. This last exhibited two sharp peaks, at frequencies $f_m$ and $2f_m$, superimposed to the broad Lorentzian shape related to the thermal motion. 

\begin{figure}[t]
\begin{center}
\includegraphics[width=0.8\linewidth]{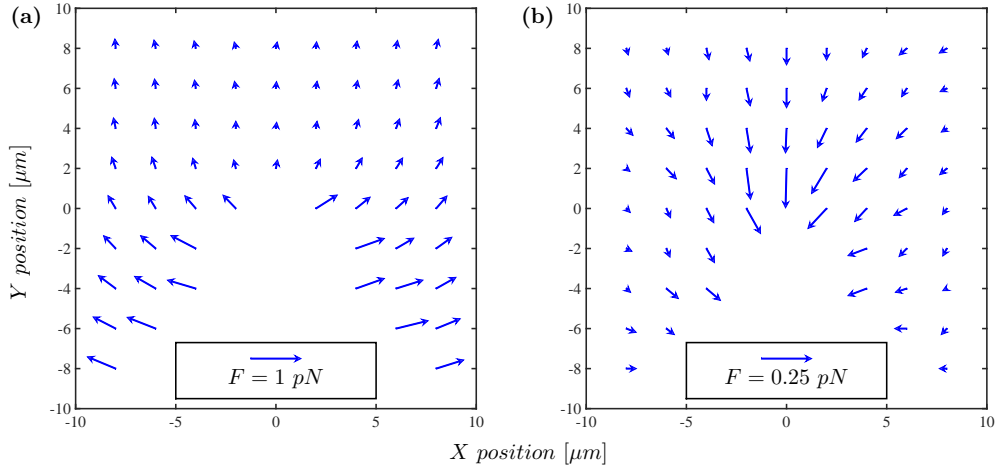}
\caption{Pattern of the electric (a) and dielectrophoretic (b) forces in  a region around the tip. The scale for DEP forces are one fourth of that for EP force.}
\label{fig:field}
\end{center}
\end{figure}

The results of the force field patterns measured in a region around the tip are shown in Fig. \ref{fig:field}. As it can be noted, both EP and DEP forces are below 1 pN, a fact which points out a high sensitivity of the technique. In particular, EP forces result about four times larger than DEP forces. In addition, at the modulation frequency used, the CM term is positive and the DEP force points toward the spatial position where the maximum electric field gradient occurs. 

Our technique can be easily generalized for the case of higher frequencies, for which the CM term is expected to change sign. Indeed, at high frequency, although the trapped bead is no more able to follow such a fast oscillation, its center displacement can be detected thanks to the high sensitivity of our position detector.

\section{Conclusion}
In conclusion, we have proved that a charged microsphere held in an optical trap can be handled as a probe for measuring simultaneously the absolute values of the electrophoretic and dielectrophoretic forces. Our method is based on the autocorrelation function, which offers high sensitivity and avoids the use of the phase-sensitive detection schemes currently employed. In particular, our approach allows one to detect very weak DEP forces. We reconstructed the pattern of  EP and DEP forces in a plane around a tip electrode, a geometry that can be generalized to more complex ones with potential 3D analysis as well. Our experiment points out the necessity to take into account the dielectrophoretic force contribution for determining the actual forces experienced by a dielectric object in presence of a high electric field gradient, and can be of potential interest for studying and mapping EP and DEP forces around scanning probe plasmonic tips that are typically characterized by highly localized electric fields\cite{RuscianoACSN14}.

\end{document}